\documentclass[noshowpacs,secnumarabic,11pt]{revtex4}
\usepackage{amssymb}
\usepackage{amsfonts}
\usepackage{amsmath}
\usepackage{graphicx}
\usepackage{psfig}
\usepackage{dcolumn}
\usepackage{bm}

\def\lessim{\lower.5ex\hbox{$\; \buildrel < \over \sim \;$}}
\def\gtrsim{\lower.5ex\hbox{$\; \buildrel > \over \sim \;$}}
\input{tcilatex}
\begin{document}

\preprint{}
\title{Thermal QCD Deconfinement Phase Transition in a Finite Volume within
the Color-Singletness Condition}
\author{M. Ladrem, A. Ait-El-Djoudi and G. Yezza}
\affiliation{Laboratoire de Physique des Particules et Physique Statistique\\
Ecole Normale Sup\'{e}rieure-Kouba,\\
B.P. 92, 16050, Vieux-Kouba, Algiers, Algeria.\\
E-mail: Ladrem@eepad.dz, Aitamel@eepad.dz }
\date{\today }

\begin{abstract}
\textbf{Abstract.} We study the finite size effects on the thermal
deconfinement phase transition from the hadronic gas phase to the QGP phase,
using a simple thermodynamic model based on the coexistence of confined and
deconfined phases in a finite volume. We first consider the bag model
partition function for the QGP, then we include the necessary color
singletness condition. By probing the behavior of some physical quantities
on the hole range of temperature, it turns out that in a finite size system,
all singularities occuring in these quantities in the thermodynamic limit $%
\left( V\longrightarrow \infty \right) $ are smoothed out, and the phase
transition is then rounded over a broadened critical region , with an
additional shifting effect of the critical point induced by the exact color
singletness requirement. An analysis of the finite size scaling behavior at
criticality of some characteristic quantities allows us to determine the
scaling critical exponents characterizing the thermal deconfinement phase
transition. The obtained results are in good agreement with those predicted
by other studies for a first-order phase transition.
\end{abstract}

\maketitle

\vspace*{0.1cm}

\section{Introduction}

It is generally believed that at sufficiently high temperatures and/or
densities a new phase of matter called the Quark Gluon Plasma (QGP) can be
created. This is logically a consequence of the quark-parton level of the
matter structure and of the dynamics of strong interactions described by the
Quantum Chromo Dynamics (QCD) theory. Its existence, however, has been
partly supported by Lattice QCD (LQCD) calculations and the Cosmological
Standard Model predictions. Experimentally, the only way to study the QCD
Deconfinement Phase Transition (DPT) occuring between the hadronic and QGP
phases is to try to recreate, in Ultra Relativistic Heavy Ion Collisions
(URHIC), conditions similar to those which occured in the early times of the
universe ($\sim 10^{-6}s$), immediately after the big bang.

If the QGP phase is (will be) created in URHIC at SPS\textbf{\ }(LHC and RHIC%
\textbf{)} colliders, the volume in which it takes (will take) place is
certainly finite. Also, in LQCD studies, the scale of\ the lattice space
volume is finite. This is our motivation to study the Finite Size Effects
(FSE) on the expected DPT from a Hadronic Gas (HG) phase to a QGP phase.
These effects are certainly important since statistical fluctuations in a
finite volume can hinder a sharp transition between the two phases. Only in
the Thermodynamic Limit (TL) are phase transitions characterized by some
singularities: $\delta -$singularities for a discontinuous PT and power-law
singularities for a continuous PT. In general, in finite systems, Finite
Size Effects (FSE) lead to a rounding of these singularities and a possible
mixing of phases. However, even in a such situation, it is possible to
obtain informations on the critical behavior. Large but finite systems show
a \textit{universal} behavior called \textquotedblleft Finite-Size
Scaling\textquotedblright\ (FSS) allowing the put of all the physical
systems undergoing a phase transition in a certain number of universality
classes. The systems in a given universality class display the same critical
behavior, meaning that certain dimensionless quantities have the same values
for all these systems. \textit{Critical exponents} are an example of these
universal quantities.

In the present work, we study the FSE on the thermal QCD DPT, using a simple
thermodynamic model based on the coexistence of the HG and QGP phases in a
finite volume, with low energy density phase and high energy density phase
equations of state \cite{SSG98}. For the HG phase, we have calculated the
Partition Function (PF) of a pionic gas as a first approximation, and for
the QGP phase, we have calculated the PF of a free gas of quarks and gluons.
This latter is first derived simply within the bag model, then we include
the exact color-singletness condition, since the color confinement
phenomenon requires that the QGP phase must have a neutral color charge.
This condition is implemented into the quantum statistical description of
the system using the Group Theoretical Projection Technique (GTPT) \cite%
{RT81}. This Color Singlet Partition Function (CSPF) can not be exactly
calculated and is usually calculated within the saddle point and gaussian
approximations in the limit $VT^{3}>>1$ \cite{EGR}. It turned out that the
use of the such obtained PF, within the Phase Coexistence Model (PCM), has
as a consequence the absence of the DPT \cite{YG02}. This is due to the fact
that the approximations used for the standard calculation of the CSPF, break
down at $VT^{3}<<1$, and this limit is attained using the PCM. We propose
then a new method for calculating a suitable CSPF, containing an expansion
in a power series of the QGP volume fraction, using the saddle point
approximation but avoiding the problem arising at $VT^{3}<<1.$ The CSPF
derived in this way allows us to accurately calculate physical quantities
describing well the DPT at finite volumes, like the order parameter, the
energy density and the entropy density, and probe the behavior of the PT at
criticality. We show that in the TL, these thermodynamic quantities exhibit
a discontinuity at a critical temperature $T_{c}(\infty ),$ which appears as
a delta function singularity in the first derivatives, i. e. in the specific
heat and the susceptibility. In a finite size system, the discontinuities
disappear and the variations of these quantities are perfectly smooth over
the hole range of temperature. The delta singularities are smeared into
finite peaks of width $\delta T(V)$, which is nothing but the broadening of
the critical region, with the maxima of the peaks occuring at an effective
transition temperature $T_{c}(V)$ shifted away from the true critical
temperature $T_{c}(\infty )$. This additional shifting effect of the
critical temperature is induced by the exact color singletness requirement.
An analysis of the FSS behavior at criticality of these maxima as well as of
the width of the transition region and the shift of the effective transition
temperature $T_{c}(V)$ relative to the true one $T_{c}(\infty )$ allows us
to determine the scaling critical exponents characterizing the thermal QCD
DPT.

\section{The QCD Deconfinement Phase Transition in a Finite Volume}

\subsection{Phase Coexistence Model and Partition Functions Calculation}

To study the QCD DPT in a finite volume, we consider a simple PCM used in 
\cite{SSG98}, in which the mixed HG-QGP phase system has a finite volume: $%
V=V_{HG}+V_{QGP}$. The parameter $\mathfrak{h}$ representing the fraction of
volume occupied by the HG: $V_{HG}=\mathfrak{h}V,$\ is then defined and can
be considered as an order parameter for the DPT. The value: $\mathfrak{h}=1$
corresponds then to a total HG phase and when $\mathfrak{h}=0,$ the system
is completely in the QGP phase. Assuming non-interacting phases
(separability of the energy spectra of the two phases), the total Partition
Function (PF) of the system can be approximatively written as:%
\begin{equation}
Z\left( \mathfrak{h}\right) =Z_{QGP}\left( \mathfrak{h}\right) Z_{HG}\left( 
\mathfrak{h}\right) .
\end{equation}%
The mean value of any thermodynamic quantity $A(T,\mu ,V)$ of the system, as
defined in \cite{SSG98}, can then be calculated by:%
\begin{equation}
\langle A(T,\mu ,V)\rangle =\frac{\int\limits_{0}^{1}A\left( \mathfrak{h}%
\right) Z\left( \mathfrak{h}\right) d\mathfrak{h}}{\int\limits_{0}^{1}Z%
\left( \mathfrak{h}\right) d\mathfrak{h}},
\end{equation}%
where $A(\mathfrak{h})$ is the total thermodynamic quantity in the state $%
\mathfrak{h}$, given in the case of an extensive quantity by: 
\begin{equation}
\mathcal{A}(\mathfrak{h})=\mathcal{A}_{HG}(T,\mathfrak{h}V)+\mathcal{A}%
_{QGP}(T,\mu ,\left( 1-\mathfrak{h}\right) V),
\end{equation}%
and in the case of an intensive quantity, by: 
\begin{equation}
\mathcal{A}(\mathfrak{h})=\mathfrak{h}\mathcal{A}_{HG}(T,\mathfrak{h}V)+(1-%
\mathfrak{h)}\mathcal{A}_{QGP}(T,\mu ,\left( 1-\mathfrak{h}\right) V),
\end{equation}%
with $A_{QGP}$ and $A_{HG}$ the thermodynamic quantities relative to the
individual QGP and HG phases, respectively.

The partition functions of both HG and QGP phases are calculated, and their
final expressions are given in the following. For a pionic gas, the PF is
simply given by:%
\begin{equation}
Z_{HG}=e^{\frac{\pi ^{2}}{30}T^{3}V_{HG}},
\end{equation}%
and for a QGP consisting of gluons and two flavors of massless quarks, with
a chemical potential $\mu $, within the bag model, the PF is given by \cite%
{AEA99}: 
\begin{equation}
Z_{QGP}=\exp \left( \left( \frac{37\pi ^{2}}{90}T^{3}+\mu ^{2}T+\frac{1}{%
2\pi ^{2}}\frac{\mu ^{4}}{T}-\frac{B}{T}\right) V_{QGP}\right) \;,
\end{equation}%
where $B$ is the bag constant, accounting for the real vacuum pressure
exercised on the perturbative vacuum.

\subsection{Finite Size Effects on the DPT}

We'll study in the following the FSE on the thermal QCD DPT, at a vanishing
chemical potential $\left( \mu =0\right) $, and for this purpose, we'll
examine the behavior of some thermodynamic quantities with temperature for
varying volume. The quantities of interest are the order parameter, which is
simply in this case the hadronic volume fraction $\mathfrak{h}\left( T,\mu
,V\right) $, the entropy density $s\left( T,\mu ,V\right) $ and the energy
density $\varepsilon \left( T,\mu ,V\right) $. Their mean values are
calculated from eq.(2), and are given by the expressions:%
\begin{equation}
\langle \mathfrak{h}\left( T,\mu ,V\right) \rangle =\frac{\left( fV-1\right)
e^{fV}+1}{fV\left( e^{fV}-1\right) }
\end{equation}%
\begin{eqnarray}
\langle s\left( T,\mu ,V\right) \rangle &=&\mathfrak{s}_{QGP}+\left( 
\mathfrak{s}_{HG}-\mathfrak{s}_{QGP}\right) \,\langle \mathfrak{h}\left(
T,\mu ,V\right) \rangle  \notag \\
\langle \varepsilon \left( T,\mu ,V\right) \rangle &=&\mathfrak{e}%
_{QGP}+\left( \mathfrak{e}_{HG}-\mathfrak{e}_{QGP}\right) \,\langle 
\mathfrak{h}\left( T,\mu ,V\right) \rangle ,
\end{eqnarray}%
with:%
\begin{equation}
\left\{ 
\begin{array}{c}
f=\left( \frac{\pi ^{2}}{30}-\frac{37\pi ^{2}}{90}\right) T^{3}-\mu ^{2}T-%
\frac{1}{2\pi ^{2}}\frac{\mu ^{4}}{T}+\frac{B}{T}\ \ \ \ \ \ \ \ \ \ \  \\ 
\mathfrak{s}_{QGP}=\frac{74\pi ^{2}}{45}T^{3}+2\mu ^{2}T;\text{ }\mathfrak{s}%
_{HG}=\frac{2\pi ^{2}}{15}T^{3}\ \ \ \ \ \ \ \ \ \ \ \ \ \ \  \\ 
\mathfrak{e}_{QGP}=B+\frac{37\pi ^{2}}{30}T^{4}+\mu ^{2}T^{2}-\frac{\mu ^{4}%
}{2\pi ^{2}};\text{ }\mathfrak{e}_{HG}=\frac{\pi ^{2}}{10}T^{4}.%
\end{array}%
\right.
\end{equation}

Fig.(1) shows the three-dimensional plots of the order parameter, the
normalized energy density $\dfrac{\langle \varepsilon \left( T,V\right)
\rangle }{T^{4}},$ and\ the normalized entropy density $\dfrac{\langle
s\left( T,V\right) \rangle }{T^{3}}$, vs temperature and system volume, at a
vanishing chemical potential $\left( \mu =0\right) $, using the common value 
$B^{1/4}=145MeV$\ for the bag constant. The first-order character of the
transition is showed by the step-like rise or sharp discontinuity of the
illustrated quantities, approaching the TL, at a critical temperature $%
T_{c}\left( \infty \right) \simeq 104.5MeV$, which reflects the existence of
a latent heat accompanying the PT. The quantities $\dfrac{\langle
\varepsilon \rangle }{T^{4}}$ and $\dfrac{\langle s\rangle }{T^{3}}$ are
traditionally interpreted as a measure of the number of effective degrees of
freedom; the temperature increase causes then a \textquotedblleft
melting\textquotedblright\ of the constituent degrees of freedom
\textquotedblleft frozen\textquotedblright\ in the hadronic state, making
the energy and entropy densities attain their plasma values. In finite
systems, the phase transition is rounded since the probability of presence
of the QGP phase below $T_{c}$, and that of the HG phase above $T_{c}$ are
finite because of the considerable thermodynamical fluctuations. The
transition region around the critical temperature $T_{c}\left( \infty
\right) $ is then broadened, acquiring a bigger width, smaller is the volume.

\section{The QCD DPT in a finite volume including the Color-Singletness
Condition}

\subsection{The Color-Singlet Partition Function of the QGP}

The partition function for a color-singlet quark-gluon plasma contained in a
volume $V_{QGP},$ at temperature $T$ and quark chemical potential $\mu ,$ is
determined using the Group Theoretical Projection Technique (GTPT)
formulated by Turko and Redlich \cite{RT81} and is given by \cite{EGR}:%
\begin{equation}
Z_{QGP}(T,\mu ,V_{QGP})=\frac{8}{3\pi ^{2}}e^{-\frac{BV_{QGP}}{T}%
}\int\limits_{-\pi }^{+\pi }\int\limits_{-\pi }^{+\pi }d\left( \tfrac{%
\varphi }{2}\right) d\left( \tfrac{\psi }{3}\right) M(\varphi ,\psi )%
\widetilde{Z}(T,\mu ,V_{QGP};\varphi ,\psi ),
\end{equation}%
where $M(\varphi ,\psi )$ is the weight function (Haar measure) given by: 
\begin{equation}
M(\varphi ,\psi )=\left( \sin \left( \tfrac{1}{2}(\psi +\tfrac{\varphi }{2}%
)\right) \sin (\tfrac{\varphi }{2})\sin \left( \tfrac{1}{2}(\psi -\tfrac{%
\varphi }{2})\right) \right) ^{2},
\end{equation}%
and $\widetilde{Z}$\ the generating function defined by:%
\begin{equation}
\widetilde{Z}(T,\mu ,V_{QGP};\varphi ,\psi )=Tr\left[ \exp \left( -\beta
\left( \widehat{H}_{0}-\mu (\widehat{N}_{q}-\widehat{N}_{\overline{q}%
})\right) +i\varphi \widehat{I}_{3}+i\psi \widehat{Y}_{8}\right) \right] ,
\end{equation}%
where $\beta =\dfrac{1}{T}$ with the units chosen as: $k_{B}=\hbar =c=1,$ $%
\widehat{H}_{0}$\ is the free quark-gluon Hamiltonian, $\widehat{N}_{q}$ $%
\left( \widehat{N}_{\overline{q}}\right) $\ denotes the (anti-) quark number
operator, and $\widehat{I}_{3}$\ and $\widehat{Y}_{8}$\ are the color
\textquotedblleft isospin\textquotedblright\ and \textquotedblleft
hypercharge\textquotedblright\ operators respectively.

The generating function $\widetilde{Z}\left( T,V_{QGP},\mu ;\varphi ,\psi
\right) $\ can be factorized into the quark contribution and the gluon
contribution as:%
\begin{equation}
\widetilde{Z}\left( T,\mu ,V_{QGP};\varphi ,\psi \right) =\widetilde{Z}%
_{quark}\left( T,\mu ,V_{QGP};\varphi ,\psi \right) \widetilde{Z}%
_{gluon}\left( T,V_{QGP};\varphi ,\psi \right) ,
\end{equation}%
where the quark contribution is given by: 
\begin{equation}
\widetilde{Z}_{quark}\left( T,\mu ,V_{QGP};\varphi ,\psi \right) =\exp \left[
\tfrac{\pi ^{2}}{12}T^{3}V_{QGP}d_{Q}\sum\limits_{q=r,g,b}\left( \tfrac{7}{30%
}-\left( \frac{\alpha _{q}-i(\frac{\mu }{T})}{\pi }\right) ^{2}+\tfrac{1}{2}%
\left( \frac{\alpha _{q}-i(\frac{\mu }{T})}{\pi }\right) ^{4}\right) \right]
,
\end{equation}%
with $q=r,\,b,\,g$ the color indices, $d_{Q}=2N_{f}$\ counts the
spin-isospin degeneracy of quarks\ and the angles$\ \alpha _{q}$ being
determined by the eigenvalues of the color charge operators in eq. (12): 
\begin{equation}
\alpha _{r}=\tfrac{\varphi }{2}+\tfrac{\psi }{3},\;\alpha _{g}=-\tfrac{%
\varphi }{2}+\tfrac{\psi }{3},\;\alpha _{b}=-\tfrac{2\psi }{3},
\end{equation}%
and the gluon contribution is given by: 
\begin{equation}
\widetilde{Z}_{gluon}\left( T,V_{QGP};\varphi ,\psi \right) =\exp \left[ 
\tfrac{\pi ^{2}}{12}T^{3}V_{QGP}d_{G}\sum\limits_{g=1}^{4}\left( -\tfrac{7}{%
30}+\left( \frac{\alpha _{g}-\pi }{\pi }\right) ^{2}-\tfrac{1}{2}\left( 
\frac{\alpha _{g}-\pi }{\pi }\right) ^{4}\right) \right] ,
\end{equation}%
with $d_{G}=2$\ the degeneracy factor of gluons and $\alpha _{g}$ $\left(
g=1,...4\right) $\ being: 
\begin{equation}
\alpha _{1}=\alpha _{r}-\alpha _{g},\;\alpha _{2}=\alpha _{g}-\alpha
_{b},\;\alpha _{3}=\alpha _{b}-\alpha _{r},\;\alpha _{4}=0.
\end{equation}%
Let us write eq. (10) on the form: 
\begin{equation}
Z_{QGP}(T,\mu ,V_{QGP})=\dfrac{8}{3\pi ^{2}}e^{\mathfrak{q}V\left(
T^{3}g_{0}(\frac{\mu }{T})-\frac{B}{T}\right) }\int\limits_{-\pi }^{+\pi
}\int\limits_{-\pi }^{+\pi }d(\tfrac{\varphi }{2})d(\tfrac{\psi }{3}%
)M(\varphi ,\psi )e^{(g(\varphi ,\psi ,\frac{\mu }{T})-g_{0}(\frac{\mu }{T}))%
\mathfrak{q}VT^{3}},
\end{equation}%
with: 
\begin{align}
g(\varphi ,\psi ,\frac{\mu }{T})& =\frac{\pi ^{2}}{12}(\frac{21}{30}d_{Q}+%
\frac{16}{15}d_{G})+\frac{\pi ^{2}}{12}\frac{d_{Q}}{2}\sum_{q=r,b,g}\left\{
-1+\left( \frac{\left( \alpha _{q}-i(\frac{\mu }{T})\right) ^{2}}{\pi ^{2}}%
-1\right) ^{2}\right\}  \notag \\
& -\frac{\pi ^{2}}{12}\frac{d_{G}}{2}\sum_{g=1}^{4}\left( \frac{\left(
\alpha _{g}-\pi \right) ^{2}}{\pi ^{2}}-1\right) ^{2},
\end{align}%
$g_{0}\left( \frac{\mu }{T}\right) $\ being the maximum of $g(\varphi ,\psi ,%
\frac{\mu }{T})$\ for $\varphi ,\psi \in \left[ -\pi ,+\pi \right] $, given
as: 
\begin{equation}
g_{0}\left( \tfrac{\mu }{T}\right) =\tfrac{\pi ^{2}}{12}(\tfrac{21}{30}d_{Q}+%
\tfrac{16}{15}d_{G})+\tfrac{\pi ^{2}d_{Q}}{12}(\tfrac{3\mu ^{2}}{\pi
^{2}T^{2}}+\tfrac{3\mu ^{4}}{2\pi ^{4}T^{4}}),
\end{equation}%
and $\mathfrak{q}$ the QGP volume fraction: $\mathfrak{q}=1-\mathfrak{h.}$

We evaluate the integral in eq. (18) by expanding $M$ to leading order in $%
\varphi $\ and $\psi ,$ and $g$\ to fourth order in $\varphi $\ and $\psi .$
Concerning the two terms appearing from $g,$\ we do an expansion of the part
depending on the volume to second order for $VT^{3}\gg 1,$ while the term
independent of the volume is expanded in a series on the form \cite{YG02}:%
\begin{equation}
e^{\mathfrak{D}\left( \varphi ,\psi \right) \mathfrak{q}}=\sum_{j=0}^{\infty
}\frac{\left( \mathfrak{D}\left( \varphi ,\psi \right) \right) ^{j}}{j!}%
\mathfrak{q}^{j}\ .
\end{equation}%
After some calculation, the final color-singlet partition function of the
QGP, $Z_{QGP},$ can then be expressed as \cite{hep-ph0207367}:%
\begin{equation}
Z_{QGP}(T,\mu ,V\mathfrak{q})\simeq \dfrac{4}{9\pi ^{2}}\dfrac{e^{\mathfrak{q%
}V\left( T^{3}g_{0}(\frac{\mu }{T})-\frac{B}{T}\right) }}{\left( a\left( 
\frac{\mu }{T}\right) VT^{3}\right) ^{4}}\left[ \sum_{n=0}^{\infty }I_{n0}%
\mathfrak{q}^{n}-\dfrac{7}{12\pi ^{2}a^{2}\left( \frac{\mu }{T}\right) VT^{3}%
}\sum_{p=0}^{\infty }I_{p1}\mathfrak{q}^{p+1}\right] ,
\end{equation}%
where the ge neral form of the integral coefficients $I_{nm}$\ is given by:%
\begin{equation}
I_{nm}=\int_{-\pi }^{\pi }\int_{-\pi }^{\pi }d\varphi d\psi
M^{(0,0)}(\varphi ,\psi )\left( n!\right) ^{-1}\left( -\frac{2}{3}\left(
\varphi ^{2}+\frac{4}{3}\psi ^{2}\right) \right) ^{n}\left( \dfrac{\varphi
^{4}}{8}+\dfrac{2\psi ^{4}}{9}+\dfrac{\varphi ^{2}\psi ^{2}}{3}\right) ^{m},
\end{equation}%
with: $a\left( \frac{\mu }{T}\right) =\left( \frac{d_{Q}}{16}\left( \frac{%
3\mu ^{2}}{\pi ^{2}T^{2}}+1\right) +\frac{3}{8}d_{G}\right) ,$ $M^{(0,0)}$\
being the expansion of $M$ to leading order in $\varphi $\ and $\psi $.

\subsection{Finite Size Effects}

The mean values of the order parameter, the energy and entropy densities can
also be computed in this case, using the obtained CSPF within the definition
(2). Their expressions for two quark flavors are respectively given by: 
\begin{equation}
<\mathfrak{h}(T,\mu ,V)>_{\csc }=1-\frac{\int\limits_{0}^{1}d\mathfrak{q}e^{%
\widetilde{f}V\mathfrak{q}}\left( \sum_{n=0}^{\infty }I_{n0}\mathfrak{q}%
^{n+1}-\dfrac{7\sum_{p=0}^{\infty }I_{p1}\mathfrak{q}^{p+2}}{12\pi
^{2}a^{2}\left( \frac{\mu }{T}\right) VT^{3}}\right) }{\int\limits_{0}^{1}d%
\mathfrak{q}e^{\widetilde{f}V\mathfrak{q}}\left( \sum_{n=0}^{\infty }I_{n0}%
\mathfrak{q}^{n}-\dfrac{7\sum_{p=0}^{\infty }I_{p1}\mathfrak{q}^{p+1}}{12\pi
^{2}a^{2}\left( \frac{\mu }{T}\right) VT^{3}}\right) }
\end{equation}%
\begin{equation}
<\varepsilon (T,\mu ,V)>_{\csc }=\mathfrak{e}_{QGP}+\mathfrak{e}_{0}+\left( 
\mathfrak{e}_{HG}-\mathfrak{e}_{QGP}\right) <\mathfrak{h}(T,\mu ,V)>_{\csc }
\end{equation}%
\begin{equation}
<s(T,\mu ,V)>_{\csc }=\mathfrak{s}_{QGP}+\mathfrak{s}_{0}+\left( \mathfrak{s}%
_{HG}-\mathfrak{s}_{QGP}\right) <\mathfrak{h}(T,\mu ,V)>_{\csc },
\end{equation}%
with:%
\begin{equation}
\left\{ 
\begin{array}{c}
\widetilde{f}(T,\mu )=\left( (\frac{\pi ^{2}}{30}+g_{0}(\frac{\mu }{T}%
))T^{3}-\frac{B}{T}\right) \ \ \ \ \ \ \ \ \ \ \ \ \ \ \ \ \ \ \  \\ 
\mathfrak{e}_{0}=-12\frac{T}{V}\ \ ;\mathfrak{s}_{0}=-\frac{12}{V}-\frac{4}{V%
}\ln \left( VT^{3}a\left( \frac{\mu }{T}\right) \right) \ ,\ \ \ \ \ 
\end{array}%
\right.
\end{equation}%
$g_{0}(\frac{\mu }{T})$ and $a\left( \frac{\mu }{T}\right) $ being taken for 
$N_{f}=2.$

Fig. (2) illustrates as in the previous section the variations of the above
quantities with temperature for different system sizes, but in this case
including the necessary condition of color-singletness for the QGP.
Additionally to the rounding of the transition in geometrically finite
systems, the color singletness requirement causes a shift of the critical
temperature to higher values, when the volume decreases. The internal
degrees of freedom being gradually \textquotedblleft
frozen\textquotedblright\ with decreasing volume, the pressure of the
color-singlet QGP, related to the energy density by the well known relation: 
$P_{QGP}=\frac{1}{3}\left( \mathfrak{e}_{QGP}-4B\right) $, is lower at a
given temperature, and the mechanical Gibbs equilibrium between the two
phases would then be reached for $T_{c}(V)>T_{c}\left( \infty \right) .$ For
small size systems, the transition is then smoothed out over a range of
temperature $\delta T\left( V\right) $, around an effective transition
temperature $T_{c}\left( V\right) $, shifted from the true transition
temperature $T_{c}\left( \infty \right) $.

\section{Finite-Size Scaling Analysis}

\subsection{Finite-Size Scaling}

In the TL, phase transitions are characterized by the appearance of
singularities in some second derivatives of the thermodynamic potential,
such as the susceptibility $\chi $ and the specific heat $c$. For a first
order phase transition, the divergences are $\delta $-function
singularities, corresponding to the discontinuities in the first derivatives
of the thermodynamic potential. For instance, in the case of the thermal QCD
DPT, the $\delta $-singularities appear in the susceptibility $\chi \left(
T,V\right) ,$ defined to be the first derivative of the order parameter\
with respect to temperature, i. e., $\chi \left( T,V\right) =\dfrac{\partial
\langle \mathfrak{h}\left( T,V\right) \rangle }{\partial T},$ and in the
specific heat density $c\left( T,V\right) $ defined as: $c\left( T,V\right) =%
\dfrac{\partial \langle \varepsilon \left( T,V\right) \rangle }{\partial T}$%
. In finite volumes, these $\delta $-functions are rounded into finite peaks
over a range of temperature $\delta T\left( V\right) $ as illustrated in
Fig. (3), and the peaks occuring at an effective transition temperature $%
T_{c}\left( V\right) $\ are shifted away from the true critical temperature $%
T_{c}\left( \infty \right) $. The width of the transition region, the shift
of the critical temperature and the maxima of the peaks of the specific heat
and the susceptibility show a scaling behavior at criticality, characterized
by scaling critical exponents as: 
\begin{equation}
\left\{ 
\begin{array}{c}
\delta T\left( V\right) \sim V^{-\theta
_{T}}\;\;\;\;\;\;\;\;\;\;\;\;\;\;\;\;\;\;\;\;\;\;\;\;\;\;\;\;\; \\ 
\tau _{T}(V)=T_{c}\left( V\right) -T_{c}\left( \infty \right) \sim
V^{-\lambda _{T}} \\ 
c_{T}^{\max }\left( V\right) \sim V^{\alpha
_{T}}\;\;\;\;\;\;\;\;\;\;\;\;\;\;\;\;\;\;\;\;\;\;\;\;\ \ \  \\ 
\chi _{T}^{\max }\left( V\right) \sim V^{\gamma
_{T}}.\;\;\;\;\;\;\;\;\;\;\;\;\;\;\;\;\;\;\;\;\;\;\;\ \ 
\end{array}%
\right.
\end{equation}

The finite size scaling (FSS) analysis is used to recover the scaling
critical exponents, and it has been shown \cite{BH88} that for a first order
phase transition, the finite size quantities $\delta T\left( V\right) $, $%
\tau _{T}(V)$ scale as: $V^{-1},$\ and the maxima $c_{T}^{\max }\left(
V\right) $\ and $\chi _{T}^{\max }\left( V\right) $\ scale as: $V$. The
critical exponents $\theta _{T}$, $\lambda _{T}$, $\gamma _{T}$ and $\alpha
_{T}$\ are then all equal to unity.

\subsection{Numerical determination of the scaling critical exponents for
the thermal QCD DPT}

\subsubsection{\textit{The susceptibility scaling critical exponent }$%
\protect\gamma _{T}$}

The plots of the susceptibility are presented in Fig.(4-left) versus
temperature for various sizes. The delta function singularity of the
susceptibility occuring in the TL is smeared, in a finite volume, into a
finite peak of width $\delta T\left( V\right) $, with the maximum of the
peak $\mid \chi _{T}\mid ^{\max }\left( V\right) $ occuring at the effective
transition temperature $T_{c}(V)$. The plot of the maxima $\mid \chi
_{T}\mid ^{\max }\left( V\right) $\ versus volume is illustrated in
Fig.(4-right), and the linearity of the data with $V$ can clearly be noted.
A numerical parametrization with the power-law form: $\mid \chi _{T}\mid
^{\max }\left( V\right) \sim V^{\gamma _{T}},$ gives the value\ of the
susceptibility critical exponent: $\gamma _{T}=0.99\pm 0.04.$ Let us note
that the associated error is a systematic one, resulting from the errors in
the localization of the maxima. This systematic error will be estimated for
all the critical exponents determined in the following.

\subsubsection{\textit{The specific heat scaling critical exponent }$\protect%
\alpha _{T}$}

The variations with temperature of the specific heat density are presented
for different sizes in Fig.(5-left), which shows the rounding of the delta
function singularity of $c\left( T,V\right) $ in finite systems into a
finite peak of width $\delta T\left( V\right) $\ and height $c_{T}^{\max
}\left( V\right) .$ For decreasing volume, the width gets larger while the
height of the peak decreases. The data of the maxima of the specific heat $%
c_{T}^{\max }\left( V\right) $\ are fitted to the power-law form: $%
c_{T}^{\max }\left( V\right) \sim V^{\alpha _{T}}$ in Fig.(5-right),\ and
the obtained specific heat critical exponent is: $\alpha _{T}=0.99\pm 0.04.$

\subsubsection{\textit{The shift scaling critical exponent }$\protect\lambda %
_{T}$}

For the study of the shift of the transition temperature $\tau
_{T}(V)=T_{c}\left( V\right) -T_{c}\left( \infty \right) $, we need to have
the finite size transition temperature\ $T_{c}\left( V\right) $. This latter
is defined to be the temperature at which the rounded peaks of the
susceptibility and the specific heat reach their maxima, and is found to be
shifted away from the true transition temperature $T_{c}\left( \infty
\right) $. Fig.(6) illustrates the results for the shift of the transition
temperature plotted versus inversed volume, and shows the linear character
of the variations. The shift critical exponent obtained from a fit to the
form: $\tau _{T}(V)\sim V^{-\lambda _{T}}$, is: $\lambda _{T}=1.0085\pm
0.0009.$

\subsubsection{\textit{The smearing scaling critical exponent }$\protect%
\theta _{T}$}

The width of the transition region can be defined by the gap: $\delta
T(V)=T_{2}(V)-T_{1}(V)$ with $T_{1}(V)$ and $T_{2}(V)$ the temperatures at
which the second derivative of the order parameter reaches its maxima, or in
other terms the temperatures at which the third derivative of the order
parameter vanishes, i. e., $\left. \frac{\partial^{3}}{\partial T^{3}}\langle%
\mathfrak{h}\left( T,V\right) \rangle\right\vert _{T_{1}(V),T_{2}(V)}=0.$

Fig.(7-left) illustrates the variations of the second derivative of the
order parameter with temperature for various sizes, and shows that the gap
between the two extrema, which represents the broadening of the transition
region, decreases with increasing volume. The results for the widths $\delta
T\left( V\right) $, plotted in Fig.(7-right) vs the inverse of the volume,
were fitted to the power law form: $\delta T\left( V\right) \sim V^{-\theta
_{T}}$, and the obtained smearing critical exponent is: $\theta
_{T}=1.016\pm 0.007.$

\section{Conclusion}

Our work has shown the influence of the finiteness of the system size on the
behavior of thermodynamical quantities near criticality. The sharp
transition observed in the thermodynamical limit, signaled by
discontinuities in the order parameter and in the energy and entropy
densities at a critical temperature $T_{c}\left( \infty \right) $, is
rounded off in finite volumes, and the variations of these thermodynamic
quantities are perfectly smooth on the hole range of temperature. All
discontinuities are rounded over a broadened critical region of width $%
\delta T\left( V\right) $\ around the critical temperature, and the delta
function singularities corresponding to these discontinuities, appearing in
the first derivatives of these quantities, i.e. in the susceptibility and
specific heat density, are then rounded into finite peaks of widths $\delta
T\left( V\right) $. The inclusion of the exact color-singletness requirement
in the QGP partition function induces an additionnal shifting effect of the
effective transition temperature $T_{c}\left( V\right) $ to higher values $%
\left( T_{c}\left( V\right) >T_{c}\left( \infty \right) \right) $. A FSS
analysis of the behavior of the width of the transition region$\ \delta
T\left( V\right) $, the shift of the effective transition temperature
relative to the true one $\tau _{T}\left( V\right) =T_{c}\left( V\right)
-T_{c}\left( \infty \right) $, and the maxima of the rounded peaks of the
susceptibility $\chi _{T}^{\max }\left( V\right) $\ and the specific heat\ $%
c_{T}^{\max }\left( V\right) $\ near criticality, shows their power-law
variations with the volume characterized by the scaling critical exponents $%
\theta _{T},\,\lambda _{T},\,\alpha _{T},\,\gamma _{T}$. Numerical results
for these critical exponents have been obtained, and the associated
systematic errors, resulting from the numerical method used for the
localization of the maxima $\chi _{T}^{\max }\left( V\right) $\ and $%
c_{T}^{\max }\left( V\right) $ and of $T_{1}(V)$ and $T_{2}(V)$, have been
estimated. These numerical values are in good agreement with our analytical
results: $\theta _{T}=\lambda _{T}=\alpha _{T}=\gamma _{T}=1$ obtained in a
parallel work \cite{LYA02}. These results are characteristic of the first
order phase transition, as predicted by different FSS theoretical approaches 
\cite{BDT2000}.

\end{document}